\documentclass[showpacs,twocolumn]{revtex4}
\usepackage{graphicx}
\usepackage{bm}
\usepackage{graphicx}
\usepackage{amsmath}
\usepackage{mathrsfs}
\usepackage{ulem}
\usepackage{color}
\usepackage[colorlinks, linkcolor=red,anchorcolor=green,citecolor=blue]{hyperref}

\begin{document}

\title{ Sequential Coalescence with Charm Conservation in High Energy Nuclear Collisions }
\author{Jiaxing Zhao$^1$, Shuzhe Shi$^2$, Nu Xu$^3$ and Pengfei Zhuang$^1$}
\address{ $^1$ Physics Department, Tsinghua University and Collaborative Innovation Center of Quantum Matter, Beijing 100084, China\\
          $^2$ Physics Department and Center for Exploration of Energy and Matter, Indiana University, Bloomington, IN 47408, USA\\
          $^3$ Institute of Modern Physics, Chinese Academy of Sciences, Lanzhou, Gansu 730000 and Institute of Particle Physics, Central China Normal University, Wuhan 430070, China}
\date{\today}

\begin{abstract}
Heavy quarks are initially produced in nuclear collisions and the number is conserved during the evolution of the system. We establish a sequential coalescence model with charm conservation and apply it to charmed hadron production at RHIC and LHC energies. The charm conservation enhances the earlier formed hadrons and reduces the later formed ones, which leads to a $D_s/D^0$ enhancement and a $\Lambda_c/D^0$ suppression. The mass dependence of the sequential hadron formation provides us a new tool for studying the quark-gluon plasma hadronization in high energy nuclear collisions.
\end{abstract}

\maketitle

There are two sources of parton production in high energy nuclear collisions, the initial production and thermal production. Since the charm quark mass $m_c\sim 1.2$ GeV is much larger than the typical temperature of the fireball formed in nuclear collisions at RHIC and LHC energies, the thermal charm quark production is weak and can be safely neglected~\cite{Zhou:2016wbo}. Therefore, the charm quark number $N_c$ is controlled only by the initial condition and conserved during the evolution of the system. Considering that $N_c$ is very small in comparison with the number of light quarks, the charm conservation is expected to strongly affect the charmed hadron production in nuclear collisions~\cite{Oh,Plumari:2017ntm,Gorenstein:2000ck}.

The question is how to realize the charm conservation in the hadronization process of the quark-gluon plasma in heavy ion collisions. With the expansion of the colliding system, the plasma temperature drops down continuously and reaches finally the critical temperature $T_c$ of the deconfinement where light quark hadronization happens. Since the hadrons containing heavy quarks are with larger binding energy, they can survive at higher temperature, namely the heavy quark hadronization happens before the light quark hadronization. This quark mass dependence of the hadronization temperature, in other words the hadron dissociation temperature, can be calculated with potential models. For instance, the $J/\psi$ dissociation temperature $T_{J/\psi}$ is much higher than that for light hadrons~\cite{Satz:2005hx,Guo:2012hx}, and the $\Upsilon$ dissociation temperature $T_\Upsilon$ is even higher~\cite{Zhou}. The hadronization sequence is also supported by the evidence of early multi-strange hadron freeze-out in high energy nuclear collisions~\cite{Hecke}. In the picture of sequential hadronization, the charm conservation effect on hadron production becomes clear: the later formed charmed hadrons are relatively suppressed with respect to the earlier produced ones. For charm-strange hadron production, the charm conservation effect will be further amplified by the strangeness enhancement~\cite{hem,Rafelski:1982pu}. Since charm-strange hadrons like $D_s^+$ are formed earlier than normal charmed hadrons like $D^0$, the enhanced $D_s$ production, due to strangeness enhancement, will naturally lead to a suppression of $D_0$. As a result, the ratio of $D_s/D^0$ is further enhanced. A significant $D_s^+/D^0$ enhancement is recently observed in heavy ion collisions at RHIC~\cite{lzhou} and LHC energies~\cite{Adam:2015jda}.

In this Letter we investigate the effect of charm conservation on hadron production in a sequential coalescence model. The coalescence mechanism is successfully used to describe hadron distributions in heavy ion collisions, like the quark number scaling law of hadron collective flow~\cite{Molnar:2003ff}, the enhancement of baryon to meson ratios~\cite{Hwa:2002tu,Fries:2003vb}, and the yield enhancement and elliptic flow of charmed hadrons~\cite{Greco:2003vf,Song:2015ykw,Lee:2007wr,He:2014tga}. We will first solve the two-body Dirac equation with lattice simulated quark-antiquark potential to determine the coalescence temperatures for charmed mesons. From the hydrodynamic equations for the evolution of the fireball, we further obtain the hadron coalescence times. We then calculate the charmed hadron spectra in the sequential coalescence model with charm conservation.

The two-body Dirac equation of constrained dynamics was successfully applied to the relativistic description of light meson spectra~\cite{Crater:2008rt} in vacuum and extended to hidden~\cite{Guo:2012hx} and open\cite{Shi:2013rga} charmed mesons at finite temperature. For a general bound state $c\bar q$ with $q=u,d,s,c$, the radial motion of the spin singlet $u_0$ and triplet $u_1^0, u_1^\pm$ relative to the center of mass is characterized by the coupled equations~\cite{Crater:2008rt,Shi:2013rga},
\begin{eqnarray}
\label{dirac}
&& (-d^2/dr^2+\Phi_V+\Phi_D+\Phi_0^0)u_0+\Phi_1^0 u_1^0 = b^2u_0,\\
&& (-d^2/dr^2+\Phi_V+\Phi_D+\Phi_0^1)u_1^0+\Phi_1^1 u_0 = b^2u_1^0,\nonumber\\
&& (-d^2/dr^2+\Phi_V+\Phi_D+\Phi_0^2)u_1^++\Phi_1^2 u_1^- = b^2u_1^+,\nonumber\\
&& (-d^2/dr^2+\Phi_V+\Phi_D+\Phi_0^3)u_1^-+\Phi_1^3 u_1^+ = b^2u_1^-,\nonumber
\end{eqnarray}
where $b^2$ as a function of meson and quark masses is the energy eigenvalue, $\Phi_V$ is controlled by the central potential $V(r)$ between $c$ and $\bar q$, $\Phi_D$ is the relativistic Darwin term, and $\Phi_0^i$ and $\Phi_1^i$ with $i=0,1,2,3$ are governed by the spin, orbital and total angular momenta. The explicit expressions for $\Phi_V$, $\Phi_D$, $\Phi_0^i$ and $\Phi_1^i$ are given in Ref.~\cite{Guo:2012hx,Crater:2008rt}. For a pair of charm quarks at finite temperature, the free energy $F_{c\bar c}$ is calculated by lattice simulation~\cite{Petreczky:2010yn} and it can be considered as the potential $V_{c\bar c}$~\cite{Burnier:2014ssa}. Taking $V(r,T)=V_{c\bar c}(r,T)$, the quark masses in the Dirac equations (\ref{dirac}) are fixed to be $m_u=m_d=0.25$ GeV, $m_s=0.35$ GeV and $m_c=1.31$ GeV by fitting the charmed meson masses in vacuum~\cite{Crater:2008rt}.

By solving the coupled dynamical equations (\ref{dirac}) we obtain the charmed meson binding energy $\epsilon(T)$ and the radial wave function $\psi(r,T)=u(r,T)/r$ from which the averaged meson size is calculated $\langle r \rangle(T)=\int dr r^3|\psi(r,T)|^2/\int dr r^2|\psi(r,T)|^2$. From the definition of hadron dissociation $\epsilon(T_h)=0$, we extract the dissociation temperatures $T_{D_s^+}=1.2\ T_c$ and $T_{D_s^{*+}}\simeq T_{D^0}\simeq T_{D^{*0}}\simeq T_{D^{*+}}=1.15\ T_c$.

The two-body Dirac equations are for charmed mesons. For charmed baryons like $\Lambda_c$, $\Sigma_c$, $\Xi_c$ and $\Omega_c$, it becomes difficult to solve the corresponding three-body Dirac equations~\cite{Whitney:2011aa} and extract the dissociation temperatures. Taking into account the relation $V_{qq}\simeq V_{q\bar q}/2$ between quark-quark and quark-antiquark potentials, it is clear that the dissociation temperature for charmed baryons should be lower than that for charmed mesons. By solving the non-relativistic three-body Schr\"odinger equation for triply charmed baryon $\Omega_{ccc}$, the dissociation temperature is only slightly above $T_c$~\cite{He:2014tga}. Since the effect of charm conservation on hadron production depends mainly on the sequence of productions, we take simply the dissociation temperature as $T_c$ for all singly charmed baryons.

\begin{figure}[htb]
{$$\includegraphics[width=0.4\textwidth]{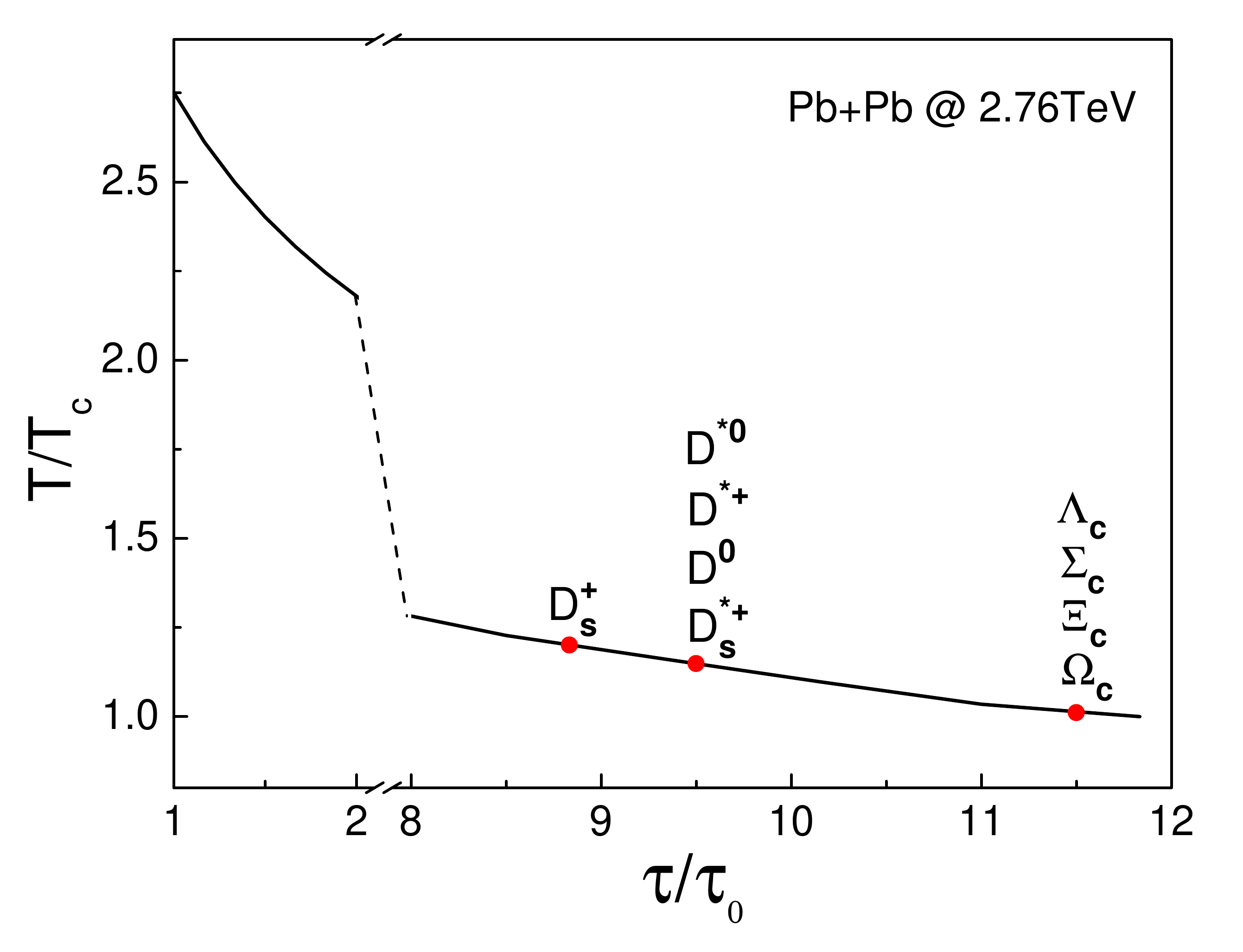}$$
\vspace{-1cm}
\caption{The time evolution of the maximum temperature in central Pb+Pb collisions at $\sqrt{s_{NN}}=2.76$ TeV. The dots indicate the coalescence temperature and time for singly charmed hadrons. $\tau_0=0.6$ fm/c and $T_c=165$ MeV are the initial time and critical temperature of the quark matter. }
\label{fig1}}
\end{figure}

The quark matter created in high energy nuclear collisions is very close to an ideal fluid and its space-time evolution is controlled by the hydrodynamic equations $\partial_\mu T^{\mu\nu}=0$ with $T^{\mu\nu}$ being the energy-momentum tensor of the system. To close the hydrodynamics, we take the equation of state of the hot medium with a first order phase transition between the ideal quark matter and hadron gas at $T_c=165$ MeV~\cite{Sollfrank:1996hd}. The initial condition at time $\tau_0=0.6$ fm/c~\cite{Hirano:2010jg} is determined by the colliding energy and nuclear geometry, which leads to a maximum initial temperature $T_0=320$ MeV in central Au+Au collisions at $\sqrt{s_{NN}}=200$ GeV~\cite{Song:2010mg} and $T_0=485$ MeV in central Pb+Pb collisions at $\sqrt{s_{NN}}=2.76$ TeV~\cite{Hirano:2010jg}. By solving the hydrodynamic equations, one obtains the local temperature $T(x)$ and fluid velocity $u_\mu(x)$. During the evolution of the quark matter, the temperature continuously drops down due to the expansion of the system. When the temperature reaches the dissociation temperature $T_h$ for a kind of charmed hadrons, they are produced via coalescence mechanism on the hyper surface $\tau_h({\bf x})$ controlled by the condition $T({\bf x},\tau_h)=T_h$. The time evolution of the temperature at the center of the fireball in centra Pb+Pb collisions at LHC energy and the corresponding coalescence times $\tau_{D_s^+}$, $\tau_{D^0}\simeq \tau_{D_s^{*+}}\simeq \tau_{D^{*0}}\simeq \tau_{D^{*+}}$ and $\tau_{\Lambda_c}\simeq\tau_{\Sigma_c}\simeq\tau_{\Xi_c}\simeq\tau_{\Omega_c}$ are shown in Fig.\ref{fig1}. It is clear to see a sequential coalescence for charmed hadrons.

We now take the sequential coalescence mechanism with charm conservation to calculate the charmed hadron spectra. The observed momentum distribution in the final state can be written as~\cite{Fries:2008hs}
\begin{eqnarray}
\label{coalescence}
\frac{dN_h}{d^2 p_Td\eta} &=& c \int p^\mu d\sigma_\mu \prod_{i=1}^n{\frac{d^4 x_i d^4 p_i}{(2\pi)^3}}f_i(x_i,p_i)\nonumber\\
&\times& W_h( x_1,...,x_i, p_1,...,p_i),
\end{eqnarray}
where the integration is on the coalescence hypersurface $\sigma_\mu(\tau_h,{\bf x})$, the product is over the constituent quarks with $n=2$ for mesons and $n=3$ for baryons, and $p_\mu =(p_0,{\bf p})$ with $p_0=\sqrt{m_{h}^2+{\bf p}^2}$ and ${\bf p}=({\bf p}_T, p_z=p_0\sinh\eta)$ is the four dimensional hadron momentum. The hadron coordinate $x$ and momentum $p$ are associated with the constituent quark coordinates $x_i$ and momenta $p_i$ via a Lorentz transformation~\cite{He:2014tga}. The constant $c$ is the statistical factor to take into account the inner quantum numbers in forming a colorless hadron, which is $1/12$ for $D^{*0}$, $D^{*+}$ and $D_s^{*+}$, $1/18$ for $\Sigma_c^*$, $1/36$ for $D_0$, $D_s^+$ and $\Sigma_c$, $1/54$ for $\Xi_c$, $\Xi_c\rq{}$ and $\Omega_c^{*0}$, and $1/108$ for $\Lambda_c$ and $\Omega_c^0$.

$W_h$ in the hadron spectra (\ref{coalescence}) is the Wigner function or coalescence probability for $n$ quarks to combine into a hadron. It is usually parameterized as a Gaussian distribution and the width is fixed by fitting the data in heavy ion collisions~\cite{Fries:2008hs}. In our calculation here, however, the Wigner function in center of mass frame for a charmed meson is directly from the wave function $\Psi(z)=\psi(r)Y(\Omega)$ where the radial part $\psi(r)$ is the solution of the Dirac equations (\ref{dirac}),
\begin{equation}
\label{wigner}
W_h(z,q)=\int d^4 y e^{-iqy}\Psi(z+y/2)\Psi^*(z-y/2).
\end{equation}
As for charmed baryons $\Lambda_c, \Sigma_c, \Xi_c$ and $\Omega_c$, we still take a Gaussian distribution~\cite{Oh} and the width is related to the mean radius of the baryon. We take $\langle r \rangle \approx 0.9$ fm~\cite{SilvestreBrac:1996bg} at the coalescence temperature for all singly charmed baryons.

$f_i$ in the spectra (\ref{coalescence}) are the distribution functions of the constituent quarks. The light quarks $u$ and $d$ are thermalized and take equilibrium distribution $f_{th}(x_i,p_i)=N_i/(e^{u_\mu  p_i^\mu/T}+1)$ with the degenerate factor $N_i=6$ and local velocity $u_\mu(x_i)$ and temperature $T(x_i)$ on the coalescence hyper surface. The strangeness enhancement, due to the thermal production in quark matter via for instance the gluon fusion process $gg\to s\bar s$, was observed in heavy ion collisions and has long been considered as a signal of the quark matter formation~\cite{Rafelski:1982pu}. Considering that strange quarks may not reach a full chemical equilibrium at RHIC energy, their thermal distribution is multiplied by a fugacity factor $\gamma_s=0.85$ at RHIC and $\gamma_s=1$ at LHC.

Initially created charm quarks interact with the hot medium and loss energy continuously. At the coalescence time, $f_c$ is in between two limits: the perturbative QCD limit without any energy loss and the equilibrium limit with full charm quark thermalization. From the experiment observation of $D$ meson elliptic flow~\cite{Adamczyk:2017xur}, the slowly moving charm quarks may have reached thermalization at the coalescence time. We take, as a first approximation, a linear combination of $f_{pp}$ and $f_{th}$ as the charm quark distribution, $f_c(x_i,p_i) = \rho_c(x_i)[\alpha f_{th}(p_i)+\beta f_{pp}(p_i)]$, where both $f_{pp}$ and $f_{th}$ are normalized distributions and $f_{pp}(p)$ can be extracted from PYTHIA simulation for p+p collisions~\cite{pythia}. The coefficients $\alpha$ and $\beta$ satisfy the constraint $\alpha + \beta =1$, and their values control the degree of thermalization of charm quarks. The perturbative limit and thermal equilibrium limit correspond respectively to $(\alpha,\beta)=(0,1)$ and $(\alpha,\beta)=(1,0)$. Considering the continuous thermalization of charm quarks and sequential coalescence of charmed hadrons, we take $(\alpha,\beta)=(0.4,0.6)$ for $D_s^+$, $(0.5,0.5)$ for $D^0,\ D_s^{*+},\ D^{*0}$ and $D^{*+}$, and $(0.6,0.4)$ for $\Lambda_c,\ \Sigma_c,\ \Xi_c$ and $\Omega_c$.

The charm quark density in coordinate space $\rho_c(x)$ is a superposition of p+p collisions~\cite{Zhou:2014kka},
\begin{equation}
\label{density}
\rho_c(x_i)=r(\tau)T_A({\bf x}_i^T)T_B({\bf x}_i^T-{\bf b}){\cosh\eta\over \tau} {d\sigma^{c\bar c}_{pp}\over d\eta},
\end{equation}
where $T_A$ and $T_B$ are thickness functions of the two colliding nuclei~\cite{Miller:2007ri}, ${\bf b}$ is the impact parameter of the collision, and the rapidity distribution of charm quarks in p+p collisions is taken as $d\sigma_{pp}^{c\bar c}/d\eta=340\ \mu b$ at RHIC energy~\cite{Adams:2004fc} and $1.2\ mb$ at LHC energy~\cite{Abelev:2012vra}. The fraction $r(\tau)$ describes the charm conservation during the sequential coalescence,
\begin{equation}
\label{conservation2}
r(\tau)=\left\{\begin{array}{ll}
1 & \tau \leq \tau_{D_s^+}\\
(N_c-N_{D_s^+})/N_c & \tau_{D_s^+} < \tau \leq \tau_{D^0}\\
(N_c-N_D)/N_c & \tau > \tau_{D^0}
\end{array}\right.
\end{equation}
where $N_D=N_{D_s^+}+N_{D^0}+N_{D_s^{*+}}+N_{D^{*0}}+N_{D^{*+}}$ is the total $D$ meson number. In usual coalescence models, all the charmed hadrons are simultaneously produced at the phase boundary, it is then difficult to distinguish the charm fractions for different charmed hadrons. To simplify the calculation, we did not consider here the coalescence for doubly and triply charmed hadrons, since their production is much weaker than the singly charmed hadrons.

To compare our calculation with the prompt data measured in heavy ion collisions, we should consider the decay from excited states to the ground state after the sequential coalescence ceases. From the experimental data in p+p collisions~\cite{pdg}, the decay contribution to $D^0$ includes $100\%$ of $D^{*0}$ and $68\%$ of $D^{*+}$, and the decay to $D_s^+$ comes only from $100\%$ of $D_s^{*+}$. For charmed baryons, all the excited states decay to the ground state. As a result, $92\%$ of $\Lambda_c$s is produced through the excited state decay~\cite{Oh}.

\begin{figure}[htb]
{$$\includegraphics[width=0.5\textwidth]{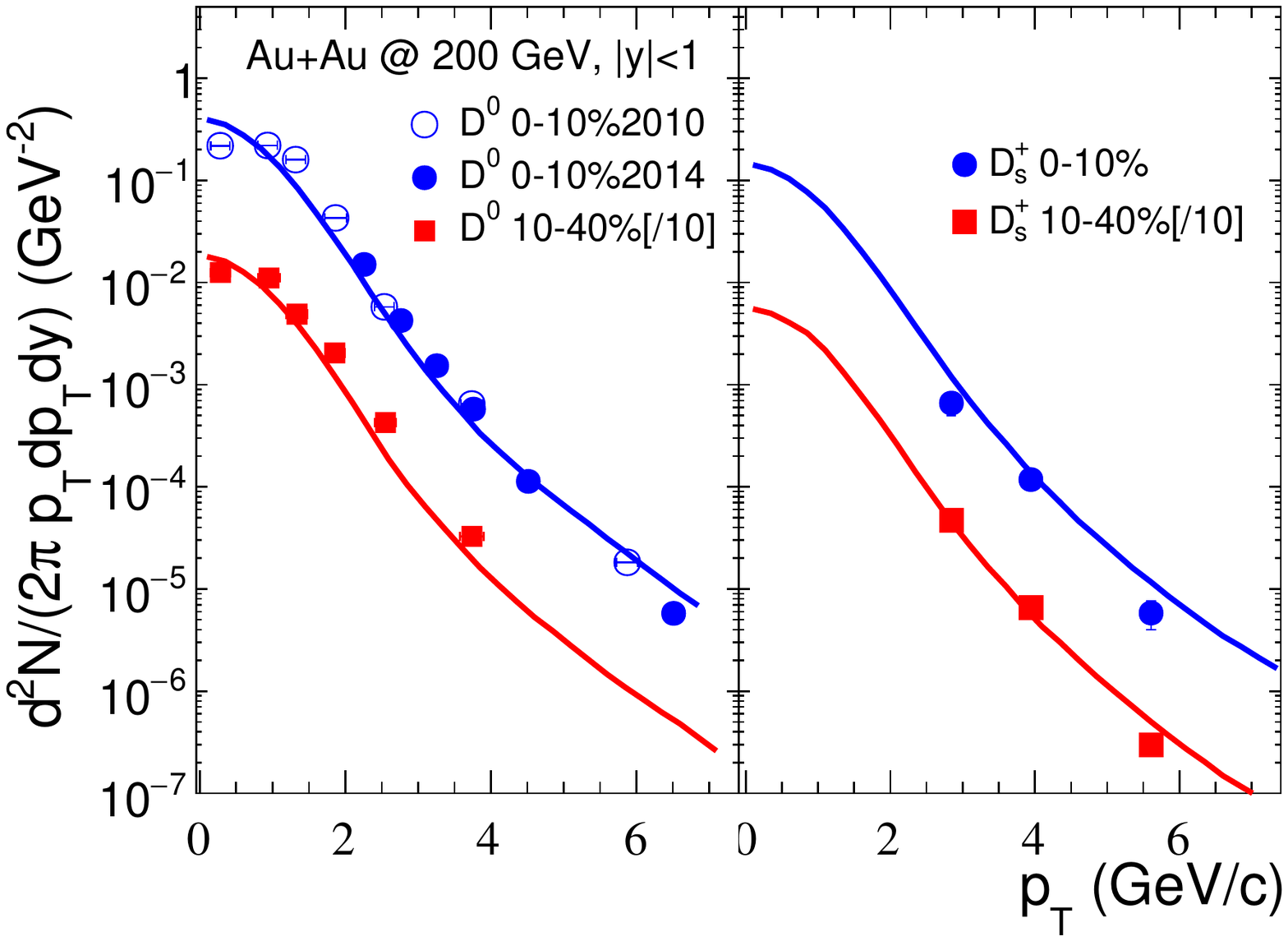}$$
\vspace{-1cm}
\caption{The transverse momentum distributions of prompt $D^0$ (left) and $D_s^+$ (right) in Au+Au collisions at RHIC energy. The lines are our sequential coalescence calculations and the data are from STAR Collaboration~\cite{Adamczyk:2014uip,Nasim:2015qbe}. }
\label{fig2}}
\end{figure}

Now we start our numerical calculation of charmed hadrons in the frame of sequential coalescence with charm conservation. The calculated transverse momentum spectra for prompt $D^0$ and $D_s^+$ and the comparison with the RHIC data are shown in Fig.\ref{fig2}. The difference in the coalescence time ($\tau_{D_s^+} < \tau_{D^0}$) and the resulted difference in the degree of thermalization of charm quarks ($\alpha=0.4$ for $D_s^+$ and $0.5$ for $D^0$) between $D_s^+$ and $D^0$ explain well the experimental data.
\vspace{-0.5cm}
\vspace{-0.5cm}
\begin{figure}[htb]
{$$\includegraphics[width=0.5\textwidth]{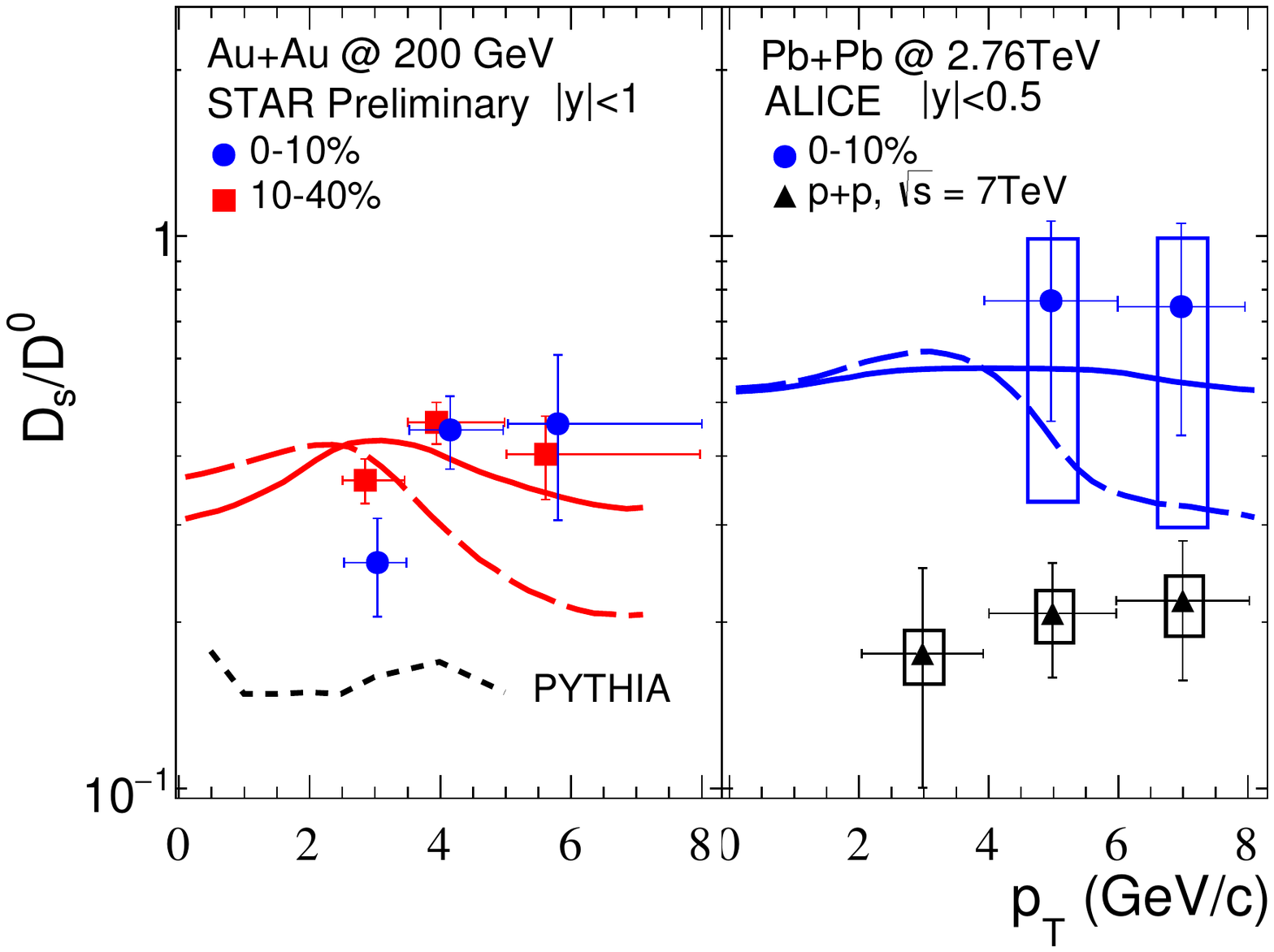}$$
\vspace{-1cm}
\caption{The yield ratio $D_s^+/D^0$ at RHIC (left) and LHC (right) energies. The experimental data in A+A and p+p collisions are from STAR~\cite{lzhou} and ALICE~\cite{Adam:2015jda, Abelev:2012tca} Collaborations, the dotted line is the PYTHIA simulation for p+p collisions~\cite{lzhou}, and the solid and dashed lines are respectively our sequential and simultaneous coalescence calculations with centrality $10-40\%$ at RHIC and $0-10\%$ at LHC. }
\label{fig3}}
\end{figure}

To see clearly the effect of charm conservation, we show in Fig.\ref{fig3} the calculated yield ratio $D_s^+/D^0$. In comparison with the p+p collisions (PYTHIA simulation at RHIC and data at LHC), the ratio in A+A collisions is significantly enhanced. The first idea coming to our mind for this enhancement is the $D_s$ enhancement due to the strangeness enhancement in quark matter. For instance, the ratio at RHIC can increase from the PYTHIA value $\sim 0.15$ to $0.2-0.3$ due to the strangeness enhancement~\cite{hem}. We calculated the ratio in two cases: One is in the sequential coalescence with charm conservation (solid lines) and the other is with a simultaneous coalescence with the charm fraction factor $r=1$ for all charmed hadrons (dashed lines). In the sequential coalescence, both the strangeness enhancement and the charm conservation are responsible for the ratio enhancement. The earlier $D_s^+$ production is with $100\%$ of charm quarks ($r=1$) but the later $D^0$ production is with only a fraction $r=(N_c-N_{D_s^+})/N_c\simeq 90\%$. On the other hand, the charm quarks involved in the later $D^0$ production are more thermalized, which leads to a shift of the $D^0$ distribution to the lower $p_T$ region in comparison with $D_s^+$. As a comprehensive result of the sequential coalescence, the ratio $D_s^+/D^0$ is strongly enhanced at intermediate $p_T$ and weakly suppressed at low $p_T$, in comparison with the simultaneous coalescence where only the strangeness enhancement is taken into account.
\vspace{-0.5cm}
\begin{figure}[htb]
{$$\includegraphics[width=0.5\textwidth]{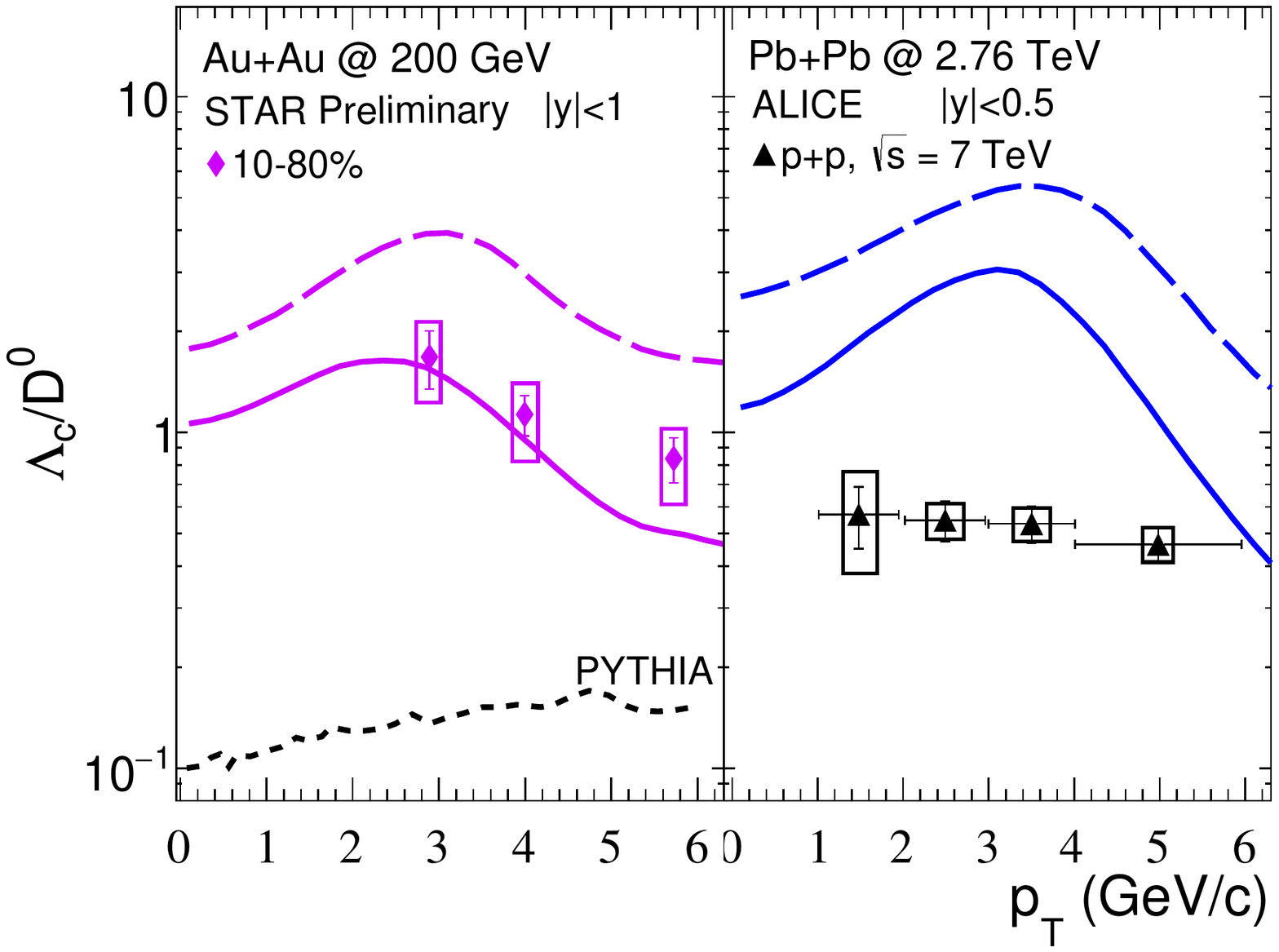}$$
\vspace{-1cm}
\caption{The yield ratio $\Lambda_c^+/D^0$ at RHIC (left) and LHC (right) energies. The experimental data in A+A and p+p collisions are from STAR~\cite{Ye_QM2018} and ALICE~\cite{Acharya:2017kfy} collaborations, the dotted line is the PYTHIA simulation for p+p collisions~\cite{Xie}, and the solid and dashed lines are respectively our sequential and simultaneous coalescence calculations with centrality $10-80\%$ at RHIC and $0-10\%$ at LHC. }
\label{fig4}}
\end{figure}

We now turn to the charmed baryon production. Fig.\ref{fig4} shows the yield ratio $\Lambda_c^+/D^0$ at RHIC and LHC energies. Considering the difference in statistics for two- and three-body states shown in the hadron spectra (\ref{coalescence}), the baryon to meson ratio in A+A collisions is dramatically enhanced in comparison with p+p collisions in coalescence models~\cite{Hwa:2002tu,Fries:2003vb}. For charmed hadron production, however, the big difference between the charm quark number fraction $r\simeq 0.6$ for $\Lambda_c$ and $r\simeq 0.9$ for $D^0$ strongly weakens the baryon to meson ratio, see the difference between the sequential coalescence with charm conservation (solid lines) and simultaneous coalescence with the same fraction $r=1$ for all the charmed hadrons (dashed lines).

The bottom quark number conservation is even better satisfied in heavy ion collisions, and the above sequential coalescence model can be directly extended to bottomed hadron production. We calculated the yield ratio $\bar B_s^0/\bar B^0$. Different from $D_s^+/D^0$ where the decay contribution from the excited states to $D^0$ is larger than that to $D_s^+$, there is no feed down from the excited states to $\bar B_s^0$ and $\bar B^0$ via strong interaction channel, and the electromagnetic channel can be experimentally separated out. Therefore, the ratio $\bar B_s^0/\bar B^0$ is clearly larger than the ratio $D_s^+/D^0$. Considering the fact that bottom quarks are more difficult to be thermalized with the medium, we use the thermalization parameters $(\alpha,\beta)=(0.3,0.7)$ for bottomed hadrons. With bottom quark mass $m_b=4.7$ GeV, we obtained from the two-body Dirac equation the dissociation temperatures $T_{\bar B_s^0}\simeq T_{D_s^+}$ and $T_{\bar B^0}\simeq T_{D^0}$ and from the sequential coalescence with bottom conservation $\bar B_s^0/\bar B^0\simeq 1$ in central Pb+Pb collisions at LHC energy .

In summary, we established a sequential coalescence model with charm conservation and applied it to charmed hadron production in high energy nuclear collisions. By solving the two-body Dirac equation for charmed mesons and hydrodynamic equations for the medium, the sequence of charmed hadron formation is determined. Mesons with charm and strange quarks are formed earlier and enhanced in yields in comparison with those with charm and light quarks. Our results on $D_s/D^0$ and $\Lambda_c/D^0$ are consistent with most recent experimental findings. The same argument applies readily to bottomed hadrons. Due to heavier masses, the estimated effect is found to be even stronger. These predictions shall be tested by future experiment data. More interestingly, our results imply that the hadron yields depend on the mass and distributions of partons understudy. The mass dependence of sequential heavy-quark hadron formation via coalescence provides a unique window for analyzing the quark-gluon plasma hadronization in high energy nuclear collisions.

\appendix {\bf Acknowledgement}: The work is supported by the NSFC and MOST grant Nos. 11335005, 11575093 and 2014CB845400.

\end{document}